\documentclass[useAMS,usenatbib,letters]{mn2e}

\usepackage{graphicx}
\usepackage{color}

\def\sss{\scriptscriptstyle}
\def\U{{\sss \!U}}
\def\L{{\sss \!L}}

\def\P{{\sss \!P}}
\def\S{{\sss \!S}}

\def\nur{\nu_\mathrm{r}}
\def\nuv{\nu_\theta}
\def\nuL{\nu_\L}
\def\nuU{\nu_\U}

\definecolor{gray}{rgb}{.6,.6,.6}
\definecolor{green}{rgb}{0,.6,0}
\definecolor{red}{rgb}{.9,0,0}

\title[Twin peak QPOs from oscillating cusp torus]{Twin peak quasi-periodic oscillations as signature of oscillating cusp torus}
\author[T\"{o}r\"{o}k et al.]{G. T\"{o}r\"{o}k$^{1}$\thanks{E-mail:
gabriel.torok@gmail.com}, K. Goluchov\'{a}$^{1}$, J. Hor\'ak$^{2}$, E. \v{S}r\'{a}mkov\'{a}$^{1}$, M. Urbanec$^{1}$,\newauthor
T. Pech\'{a}\v{c}ek$^{1}$, and P. Bakala$^{1}$\\
$^{1}$Institute of Physics, Faculty of Philosophy \& Science,
        Silesian University in Opava, \\
        Bezru\v{c}ovo n\'am.~13, CZ-746\,01 Opava,
        Czech Republic\\
$^{2}$Astronomical Institute, 
        Bo\v{c}n\'{i} II 1401/2a, 
        CZ-14131 Praha 4 - Spo\v{r}ilov,
        Czech Republic}
\begin{document}

\date{}

\pagerange{\pageref{firstpage}--\pageref{lastpage}} \pubyear{2015}

\maketitle

\label{firstpage}

\begin{abstract}
Serious theoretical effort has been devoted to explain the observed frequencies of twin-peak quasi-periodic oscillations (HF QPOs) observed in low-mass X-ray neutron star binaries. Here we propose a new model of HF QPOs. Within its framework we consider an oscillating torus with cusp that changes location $r_0$ of its centre around radii very close to innermost stable circular orbit. The observed variability is assigned to global modes of accreted fluid motion that may give strong modulation of both accretion disc radiation and the accretion rate. For a given spacetime geometry, the model predicts that QPO frequencies are function of single parameter $r_0$. We illustrate that the model can provide fits of data comparable to those reached by other models, or even better. In particular it is compared to relativistic precession model. Moreover, we also illustrate that the model consideration is compatible with consideration of models of a rotating neutron star in the atoll source 4U~1636-53.
\end{abstract}


\section{Introduction}

Many models have been proposed to explain a phenomenon of high-frequency quasi-periodic oscillations observed in neutron-star low-mass X-ray binaries (HF QPOs in LMXBs). It is believed that HF QPOs are carrying signatures of strong gravity and dense matter composition. Serious theoretical effort has been devoted to explain the observed frequencies and their correlations \citep[see, e.g.,][for further informations and a large list of references to various individual models of HF QPOs]{Tor-etal:2012}. In this Section we only briefly recall two particular outstanding theoretical frameworks.  

One of the first QPO models, the so called relativistic-precession model (RP model) identifies the twin-peak kHz QPO frequencies $\nuU$ and $\nuL$  with two fundamental frequencies of a nearly circular geodesic motion: the Keplerian orbital frequency and the periastron-precession frequency,
\begin{equation}
\nuU=\nu_\mathrm{K},\quad \nuL=\nu_\mathrm{per}=\nu_\mathrm{K} - \nu_r\,,
\end{equation}
where $\nu_r$ denotes the radial epicyclic frequency. The correlations among them is then obtained by varying the radius of the underlying circular orbit in a reasonable range. Within this framework it is usually assumed that the variable component of the observed X-ray signal originates in a bright localized spot or blob orbiting the neutron star on a slightly eccentric orbit. The observed radiation is then periodically modulated due to the relativistic effects. It has been shown that the model is roughly matching the observed $\nuU(\nuL)$ correlations \citep{Ste-Vi:1999, bel-etal:2007a, Tor-etal:2012}. Nevertheless the RP model also suffers some theoretical difficulties. It is not clear whether the modulation of a radiation from a small localized spot can produce sufficiently strong signal modulation to explain a relatively large observed HF QPO amplitudes. It is then expected that larger spots (giving higher amount of modulated photons) can undergo a serious shearing due to the differential rotation in the surrounding accretion disc. This does not agree with a high coherence of the HF QPO signal which is often observed. The model also lacks an explanation of inferred existence of preferred orbits which should be responsible for appearance of HF QPO pairs (twin peaks) and clustering of their frequencies.       

Only slightly later, \cite{abr-klu:2001} and \cite{klu-abr:2001b} proposed the concept of orbital resonance models. Within this concept, HF QPOs originate in resonances between oscillation modes of the accreted fluid. The most quoted, so-called 3:2 epicyclic resonance model, identifies the resonant eigenfrequencies with frequencies $\nuv$ and $\nur$ of radial and vertical epicyclic axisymmetric modes of disc (or torus) oscillations. It is assumed that
\begin{equation}
\nuU=\nuv,~\nuL=\nur \Leftrightarrow \nuU/\nuL=3/2\,,
\end{equation}
while the correlation $\nuU(\nuL)$ arises from resonant corrections to the eigenfrequencies \citep{abr-etal:2005:AN}. 
We stress that the model deals with a collective motion of the accreted matter.\footnote{A different class of models dealing with collective motion of accreted matter considers normal modes of accretion disc oscillations, referred to as discoseismology \citep[e.g.][]{wag-etal:2001,kat:2001}.} Moreover, the oscillation modes of innermost region of the accretion flow can modulate the amount of matter transferred to NS surface through the boundary layer \citep{pac:1987,abr-etal:2007,hor:2005}. Therefore, it may naturally explain both high amplitudes and coherence of the HF QPOs. Nevertheless, it is questionable whether the resonant corrections to the eigenfrequencies can be large enough to explain the whole observed range of $\nuU$ and $\nuL$. Furthermore, it has been shown that the model implies large range of NS masses and has difficulties when confronted to models of rotating NS based on up-to-date equations of state \citep[EoS,~see][]{urb-etal:2010:AA,Tor-etal:2012}.

Motivated by partial success of the above models and their complementary difficulties, we present a modified framework for interpreting the HF QPOs. 

\section{Oscillating cusp tori} \label{sec:model}
Our model is largely based on the theoretical work of \cite{str-sram:2009}. We adopt Kerr geometry to describe slowly rotating compact NS. We assume that the innermost region of accretion flow is hot enough to form a  pressure supported torus of a moderate thickness. Assuming a non-relativistic polytropic equation of state and neglecting the poloidal components of the fluid velocity (so that the fluid follows circular orbits), the equilibrium torus shape and its structure are completely determined by the Lane-Embden function, which is given by a simple analytic formula \citep{str-sram:2009, abr-etal:2006}
\begin{equation}
	f = 1 - \frac{1}{n c_{\mathrm{s}0}^{2}}\ln\frac{\mathcal{E}}{\mathcal{E}_0}.
	\label{eq:Lane-Embden}
\end{equation}
In this equation, $\mathcal{E}$ denotes the energy of a particle on a (non-geodesic) circular orbit having the specific angular momentum $\ell$. We assume that the angular momentum is constant in the whole volume of the torus, $\ell=\ell_0=\mathrm{const}$. Since we assume that the torus is located in the vicinity of the innermost stable circular orbit (ISCO, $r=r_{\mathrm{ms}}$) where also Keplerian angular momentum is nearly constant, we believe it is a reasonable approximation. Meaning of other symbols in eq.~(\ref{eq:Lane-Embden}) is straightforward: $n$ is the polytropic index ($n=3$ for a radiation pressure dominated fluid) and $c_{\mathrm{s}0}$ is the sound speed at the center of the torus located at radius $r_0$ in the equatorial plane, where the pressure gradient vanishes and energy $\mathcal{E}$ takes the value $\mathcal{E}_0$. Vanishing of the pressure forces at the torus center implies the streamline $r=r_0$, $\theta=\pi/2$ to be a geodesic line because of which the fluid angular momentum takes Keplerian value at that radius, $\ell_0 = \ell_\mathrm{K}(r_0)$. 

\subsection{Torus size}

The surfaces of constant density and pressure coincide with those of constant $f$ and their values can be calculated from $f$ by $\rho=\rho_0 f^n$ and $p=p_0 f^{n+1}$, where $\rho_0$ and $p_0$ refer to the values at the torus center that corresponds to $f=1$. On the other hand, the surface of the torus, where both pressure and density vanishes, is given by the condition $f=0$. It is also worth to note that the position of the center $r_0$ and the shape of these surfaces are entirely given by the value of $\ell_0$ and the spacetime geometry, while the particular values of $p$ and $\rho$ and consequently the location of the overall surface of the torus are set by the central value of the sound speed $c_{\mathrm{s}0}$. 

\citet{str-sram:2009} introduce a dimesionless parameter $\beta$ that characterizes the size of the torus,
\begin{equation}
	\beta = \frac{\sqrt{2n}c_{\mathrm{s}0}}{{r_0\mathcal{E}_0(\ell_0 g^{\phi\phi}_0 - g^{t\phi}_0})}.
	\label{eq:beta}
\end{equation}
This parameter is roughly proportional to the Mach number of the flow at the torus center as can be seen from its Newtonian limit $\beta = \sqrt{2n}(c_\mathrm{s}/r\Omega)_0$ \citep[see also][]{bla:1985}. In addition, it is also roughly proportional to the ratio of the radial (or vertical) extension of the torus to its central radius $r_0$. Hence, the sound-crossing time and the dynamical timescale of the torus are roughly similar.

\subsection{Marginally overflowing tori (cusp tori)}

The stationary solution does not exist for an arbitrary large value of $\beta$ \citep{abr-jar-sik:1978}. In addition to the obvious limit $\beta\leq\beta_\infty$ corresponding to tori whose outer edge extends to infinity, there is a much stronger constrain coming from general relativity that significantly limits the torus size close to ISCO. Large enough tori that extend below the ISCO radius, may be terminated there by a ``cusp'', where the rotation of the flow becomes Keplerian again. This is a consequence of the fact that the Keplerian angular momentum reaches its minimum at ISCO. 

The cusp corresponds to a saddle point of the Lane-Embden function and the corresponding self-crossing equipotential limits the surface of any stationary rotating fluid configuration with given angular momentum $\ell_0$. Fluid that appear outside this surface, is accreted onto the central star on the dynamical timescale driven by gravity and pressure forces without need of viscosity \citep{pac:1977}. \citet{abr-jar-sik:1978} calculated analytically the accretion rate from a slightly overflowing torus, their result agrees very well with numerical simulations. 

The critical value of $\beta$ giving a marginally overflowing torus follows from equations (\ref{eq:Lane-Embden}) and (\ref{eq:beta}),
\begin{equation}
	\beta_\mathrm{c}(r_0) = \frac{\sqrt{2\ln\left(\mathcal{E}_\mathrm{c}/\mathcal{E}_0\right)}}{r_0\mathcal{E}_0(\ell_0 g^{\phi\phi}_0 - g^{t\phi}_0)},
\end{equation}
where $\mathcal{E}_\mathrm{c}=\mathcal{E}(r_\mathrm{c})$ is the particle energy at the cusp. Its location $r=r_\mathrm{c}$ can be found by equating the Keplerian angular momentum to the fluid angular momentum $\ell_0$. 
This procedure leads to the third-order algebraic equation (in $\sqrt{r_\mathrm{c}}$), giving the position of the cusp in terms of the location of the torus center. 

If the stellar spin is neglected ($j=0$), the equation is reduced to the quadratic one and its solution can be expressed as
\begin{equation}
	r_\mathrm{c} = r_0\left(\frac{M+\sqrt{(2r_0-3M)M}}{r_0-2M}\right)^2,
	\quad
	r_0\geq 6M
\end{equation}
and the critical $\beta$-parameter reads 
\begin{equation}
	\beta_\mathrm{c} = \frac{(r_0-r_\mathrm{c})(r_0-2M)^2[r_0 r_\mathrm{c} - 2M(r_0+2r_\mathrm{c})]^{1/2}}{r_\mathrm{c} r_0(r_\mathrm{c}-2M)^{1/2}(r_0-3M)^{1/2}}.
\end{equation}
This can be used for $r_0 \la 10.47 M$ where $\beta=\beta_\infty$.

\subsection{Frequencies of epicyclic oscillations}

\citet{abr-etal:2006} pointed out the existence of the radial and vertical epicyclic modes that describes a global motion of the torus. They have found that, in the limit of infinitesimally slender tori $\beta\rightarrow 0$, frequencies of these modes $\nu_\mathrm{R}$ and $\nu_\mathrm{V}$ measured in the fluid reference frame coincide with the epicyclic frequencies of test particles,
\begin{eqnarray}
	\nu_r &=& \left(1 - \frac{6M}{r} + \frac{8jM^{3/2}}{r^{3/2}} - \frac{3j^2M^2}{r^2}\right)^{1/2}\nu_\mathrm{K}, \\
	\nu_\theta &=& \left(1 - \frac{4jM^{3/2}}{r^{3/2}} + \frac{3j^2M^2}{r^2}\right)^{1/2}\nu_\mathrm{K},
\end{eqnarray}
while at fixed azimuth their frequencies are given by $\nu_{\mathrm{R},m} = \nu_r + m\nu_\mathrm{K}$ and $\nu_{\mathrm{V},m} = \nu_\theta + m\nu_\mathrm{K}$ with $m$ being the integer azimuthal wavenumber. In particular, the $m=-1$ radial and vertical modes give the frequencies of the periastron and nodal precession of a weakly eccentric and tilted torus. It is also worth to note that they describe a collective motion of the fluid, rather then a motion of individual particles.  

In a more realistic case, when $\beta> 0$, the pressure gradient contributes to the restoring force of the perturbed torus shifting their frequencies to new `corrected' values,  
\begin{eqnarray}
	\nu_{\mathrm{R},m}(r_0,\beta) &=& \nu_r(r_0) + m\nu_\mathrm{K}(r_0) + \Delta\nu_{\mathrm{R},m}(r_0,\beta),
	\\
	\nu_{\mathrm{V},m}(r_0,\beta) &=& \nu_\theta(r_0) + m\nu_\mathrm{K}(r_0) + \Delta\nu_{\mathrm{V},m}(r_0,\beta).
\end{eqnarray}
The pressure corrections $\Delta\nu_{\mathrm{R},m}$ and $\Delta\nu_{\mathrm{R},m}$ have been calculated by \citet{str-sram:2009} using perturbation expansion in $\beta$-parameter. They found that the first nonzero corrections are of the order of $\beta^2$.

\begin{figure*}
\begin{center}
a)\hfill b)\hfill c)\hfill\,\\
\includegraphics[width=0.95\hsize]{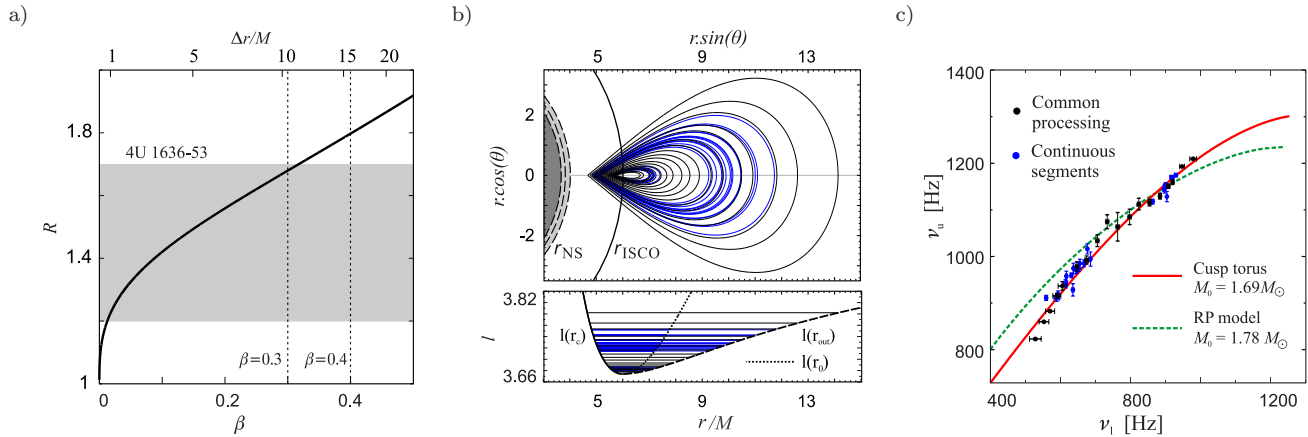}
\end{center}
\caption{a) Relation between the thickness of cusp torus and the expected frequency ratio $R$. The vertical shadow region indicates the interval of $R$ corresponding to most of the available data of the atoll source 4U~1636-53. b) Sequence of cusp tori corresponding to a one-parametric fit 
($j=0$). The color-coding is the same as in panel c). Neutron star radii $r_{\mathrm{NS}}$ are drawn for three particular NS EoS (Gle, APR, and GAN) that are further assumed within Figure~\ref{figure:2}. The bottom panel indicates angular momentum behaviour together with positions of the torus centre $r_{\mathrm{0}}$ and both inner and outer edge $r_{\mathrm{c}}$ and $r_{\mathrm{out}}$. c) Corresponding frequency relation plotted together with the datapoints. For the sake of comparison we also present the best fit implied by the RP model $(j=0)$.}
\label{figure:1}
\end{figure*}

\section{Frequency identification}
We identify the observed HF QPO frequencies with frequencies of the epicyclic modes of torus oscillations. We propose that the upper kilohertz QPO frequency is the Keplerian orbital frequency of the fluid at the center of the torus, where both pressure and density peaks and from which most of  the torus radiation emerges. The lower kilohertz QPO corresponds to the frequency of the non-axisymmetric $m=-1$ radial epicyclic mode. Overall, there is
\begin{equation}
	\nuU \equiv \nu_\mathrm{K}(r_0), \quad
	\nuL \equiv \nu_{\mathrm{R},-1}(r_0,\beta).	
\end{equation}
The QPO frequencies are then strong functions of the position of the center of the torus $r_0$ and its thickness $\beta$. Obviously, the choice of $\beta=0$ (slender tori) recovers the RP model frequencies. In the case of a finite thickness $\beta>0$, they also weakly depend on the value of the polytropic index $n$. In next, we fix $n=3$ as the inner parts of the accretion flow are believed to be radiation-pressure dominated.

We assume the cusp configuration
\begin{equation}
\label{equation:cusp}
	\beta(r_0) \doteq \beta_\mathrm{c}(r_0).
\end{equation}    
In other words, we expect that for given $r_0$ the torus is always close to its maximal possible size, just filling its `Roche-like' lobe. Thus, for a given accreting central compact object, our model predicts that the QPO frequencies are functions of a single parameter $r_0$,
\begin{equation}
\label{equation:correlation}
	\nu_\mathrm{u} \equiv \nu_\mathrm{K}(r_0), \quad
	\nu_\mathrm{l} \equiv \nu_{\mathrm{R},-1}\left[r_0,\beta_\mathrm{c}(r_0)\right]\,. \quad
\end{equation}
Therefore, one obtains a unique correlation among them by changing this parameter in a physically reasonable range. 

\subsection{Applicability of the adopted approximation}

The above equations are exact for Kerr spacetimes and represent an acceptable approximation for high-mass neutron stars \citep{urb-etal:2013}. Another restriction on their applicability follows from the adopted description of torus dynamics assuming a second order expansion in $\beta_\mathrm{c}$. Consequently we can assume only $\beta_\mathrm{c}\la 0.3$. For $j=0$, this corresponds to tori with radial extension $r_{\mathrm{out}}-r_{\mathrm{c}}\equiv\Delta r\la 10M$.  Neglecting the effects of neutron star rotation, equation (\ref{equation:correlation}) implies a relation between torus thickness $\beta_\mathrm{c}$  and frequency ratio $R\equiv\nu_\mathrm{u}/\nu_\mathrm{l}$. We show this relation in Figure~\ref{figure:1}a. From this figure we can see that our approximation can be well applied when $R\la 1.7$, while for $R>2$ it is not sufficient.

\section{Application to observed Twin Peak QPOs} \label{sec:application}

\cite{lin-etal:2011} and \cite{Tor-etal:2012} have confronted several HF QPO models with data of the atoll source 4U 1636-53 displaying twin peak QPOs mostly within the range of $R\in[1.25-1.7]$. They have outlined comparison between individual matches of the model to the data as well as quantitative estimates of the inferred NS parameters. The datapoints assumed in the former study come from a common processing of a large amount of data while the datapoints used in the latter study correspond to individual continuous observations of the source. As discussed in \cite{Tor-etal:2012}, results of both studies are consistent.

Here we confront the cusp torus model with the previously examined data. We primarily assume the datapoints of \cite{lin-etal:2011} that span a larger range of frequencies but we also check the results for the datapoints corresponding to continuous segments of observation.

\subsection{Non-rotating approximation}

First, we investigate the case of a simple one-parametric fit assuming non-rotating NS. This way we can obtain comparison with the RP model and a rough estimate of the NS mass implied by our cusp torus model.
In Figure~\ref{figure:1}b we plot the sequence of equipotential contours of cusp tori that provide the best match to the 4U~1636-53 data. In Figure~\ref{figure:1}c we show this best fit. The best fit of the RP model is included for the sake of comparison. Clearly, the cusp torus model matches the observed trend better than the RP model. We have {$\chi^2/\mathrm{dof}=2.3$} for the torus model while it is {$\chi^2/\mathrm{dof}=16.4$} for the RP model. The NS mass inferred from the cusp torus model, within $2\sigma$ confidence level, reads
\begin{equation}
M_0= 1.69 [\pm {0.02}]\,M_{\odot}\,.
\label{equation:M:nonrotating}
\end{equation}
For the datapoints corresponding to continuous segments of observation we obtain the same mass, $M_0=1.69[\pm {0.01}]\,M_{\odot}$.

\subsection{Consideration of NS rotation}

In analogy to results of \cite{Tor-etal:2012}, we may expect that the mass (\ref{equation:M:nonrotating}) belongs to a mass--angular-momentum relation implied by the cusp torus model. The result of the two-dimensional fitting of the parameters $M$ and $j$ is shown in Figure~\ref{figure:2}a.  Indeed, the best fits are reached when $M$ and $j$ are related through the specific relation. This relation can be approximated by a quadratic form, 
$M = M_0(1 + 0.68(j + j^2)),$
which results for each of the two datasets.

\section{Discussion and conclusions}\label{conclus}

There is good evidence on the NS spin frequency of 4U~1636--53 based on X-ray burst measurements. Depending on the hot-spot model consideration, the spin $\nu_\S$ reads either $\nu_\S\doteq290$\,Hz or $\nu_\S\doteq580$\,Hz \citep{stro-Mar:2002}. The value of 580Hz is usually preferred. We can therefore roughly compare the cusp torus model predictions to predictions of models of rotating NS.

\subsection{Neutron star equation of state and radius}

\begin{figure*}
\begin{center}
a)\hfill b)\hfill ~~~~~~~~~~~~~~~~~~~~~~~~~~~~~~~~\\
\includegraphics[width=0.95\hsize]{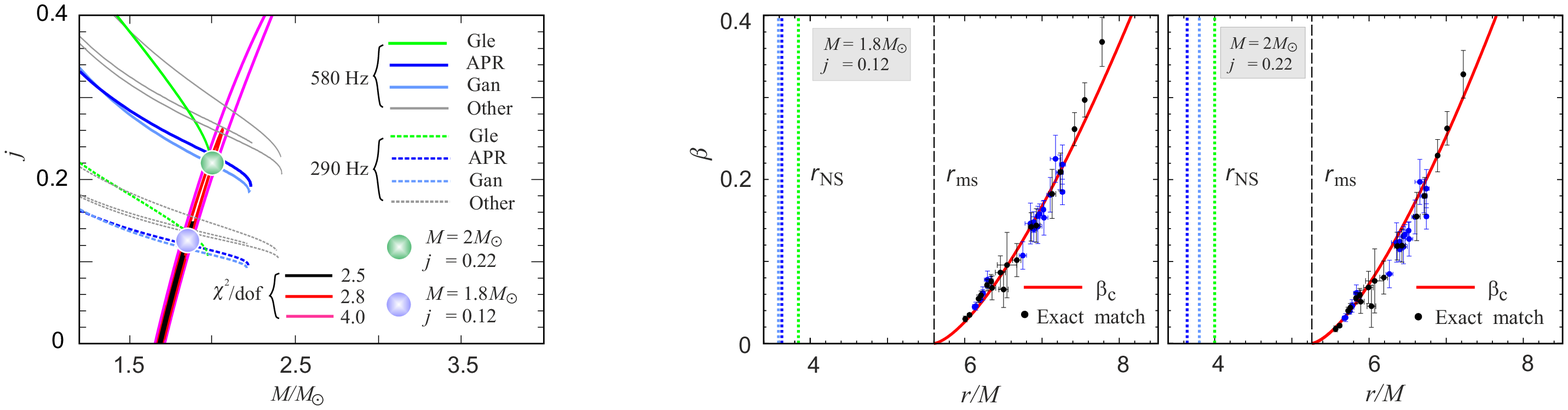}
\end{center}
\caption{a) The mass-angular momentum countours resulting from fitting of datapoints by the cusp torus model vs. mass-angular momentum relations predicted by models of rotating NS. These are drawn for several NS EoS and spin 290Hz or 580Hz inferred from the X-ray burst measurements. The two spots indicate chosen combinations of angular momentum  where the QPO model and EoS relations overlap. b) Consideration of combinations of $\beta$ and $r$ exactly matching individual datapoints for the two chosen combinations of mass and angular momentum. The color-coding is the same as in Figure~\ref{figure:1}c. The red line denotes the numerically calculated cusp torus relation.} 
\label{figure:2}
\end{figure*}

In Figure~\ref{figure:2}a we include several mass-angular momentum relations expected from models of rotating NS. We assume the following set of EoS - {SLy 4}, {APR}, {AU-WFF1, UU-WFF2 and WS-WFF3} \citep{rik-etal:2003, akm-etal:1998,wir-etal:1988,ste-fri:1995}.\footnote{In our calculations we follow the approach of \cite{har-tho:1968}, \cite{cha-mil:1974}, \cite{mil:1977}, and \cite{ urb-etal:2013}.}

Inspecting Figure~\ref{figure:2}a we can see that there are overlaps between the relations given by models of rotating stars and the relation inferred from the cusp torus model. In the figure we denote two particular values of angular momentum together with corresponding masses that roughly represent these overlaps. Assuming the two chosen combinations of mass and angular momentum, we attempted in Figure~\ref{figure:2}b to fit the data by the torus frequencies considering \emph{any} torus thickness, not only $\beta_\mathrm{c}$. We searched  for the combinations of $\beta$ and $r$ exactly matching each individual datapoint. Clearly, the obtained values are distributed very close to the cusp relation (\ref{equation:cusp}), $\beta=\beta_\mathrm{c}(r)$, where we have $r>r_{\mathrm{NS}}$.

\subsection{Model perspectives}

A more careful and computationaly demanding investigation of the spin influence following the work of \cite{str-sram:2009} and assuming the Hartle-Thorne spacetime should be applied in a consequent work following our study assuming high mass NS approximation. Nevertheless, we can conclude that there is a very strong indication that twin peak QPOs can be identified with a particular non-axisymmetric $m=-1$ radial epicyclic mode and Keplerian orbital motion associated to the cusp torus. These modes may naturally give strong modulation of both emerging radiation and the accretion rate and their  eigenfunctions change only weakly on the spatial scale of the turbulent motion. They are therefore very good candidates for explaining high amplitudes of HF QPOs. 

\subsubsection{Low frequency QPOs}

The presented concept has the potential to explain also the observed low frequency QPOs. As noticed by \cite{Klu-Ros:2013} and \cite{Ros-Klu:2014} the frequencies of vertical modes seem to be very sensitive to the NS quadrupole moment. Their consideration thus may exceed the framework of Kerr spacetime approximation adopted here. However, we roughly investigated also the frequencies of non-axisymmetric $m = −1$ vertical epicyclic mode of cusp tori. This mode corresponds to a low-frequency global precession of inclined torus and is analogical to the ``tilted hot flow precession'' discussed by \citet{Ingram+Done2010}.  Assuming the same mass, angular momentum and radii as those in Figure~\ref{figure:2} we obtained values of tens of Hertz that are of the same order as the observed frequencies. The $m=-1$ vertical epicyclic mode may therefore play the same role in the framework of cusp torus model as the Lense-Thirring preccession in the framework of RP model.

\vspace{-0.4cm}

\section*{Acknowledgments}

GT, ES, and MU would like to acknowledge the Czech grant GA\v{C}R 209/12/P740. JH was supported by the grant LH14049. We also acknowledge the internal grants of SU Opava, SGS/11/2013 and IGS/12/2015. We thank to the anonymous referee for his/her comments and suggestions that have significantly helped to improve the paper. We are grateful to Marek Abramowicz, Wlodek Kluzniak (CAMK), John Miller (University of Oxford), and Luigi Stella (INAF) for many useful discussions. Furthermore we would like to acknowledge the hospitality of the University of Oxford and the Astronomical Observatory in Rome.

\label{lastpage}
\end{document}